\documentclass[grl]{agu2001}
\usepackage{amsmath}
\usepackage{amssymb}
\usepackage{graphicx}
\usepackage{bm}
\usepackage{url}

\journalid{}
\articleid{}{}
\paperid{}
\cpright{}{}
\received{} \revised{} \accepted{} \published{}

\authorrunninghead{SWISDAK AND DRAKE}
\titlerunninghead{X-line Orientation}
\authoraddr{M. Swisdak,
IREAP, University of Maryland, College Park, MD 20742-3511, USA,
(swisdak@umd.edu)}

\begin{document}
\title{The Orientation of the Reconnection X-line}

\authors{M. Swisdak \altaffilmark{1} and
J. F. Drake\altaffilmark{1,2} }

\altaffiltext{1}
{IREAP, University of Maryland, College Park, MD 20742-3511, USA}

\altaffiltext{2}
{SSL, University of California, Berkeley, CA 94720-7450, USA}

\begin{abstract}

We propose a criterion for identifying the orientation of the X-line
when two regions of plasma with arbitrary densities, temperatures, and
magnetic fields undergo reconnection.  The X-line points in the
direction that maximizes the (suitably-defined) Alfv\'en speed
characterizing the reconnection outflow.  For many
situations a good approximation is that the X-line bisects the angle
formed by the magnetic fields.

\end{abstract}

\begin{article}

\section{Introduction}

Reconnection is the dominant process by which energy is transferred
from the magnetic field to the thermal and bulk motions of the
particles in collisionless plasmas such as the magnetosphere and the
solar corona.  Both theoretical models and numerical simulations of
reconnection usually consider highly symmetric cases, e.g., the
merging of two plasmas that are identical except for their
anti-parallel fields, where symmetry considerations dictate the
reconnection plane and the orientation of the X-line (the normal
to that plane).  Realistic configurations are often more complex, as
for instance at the magnetopause where a low-density, strong-field
plasma (the magnetosphere) merges at an arbitrary angle with a
high-density, weak-field plasma (the magnetosheath).

\cite{sonnerup74a} argued that in such complex systems the orientation
of the X-line is fixed by requiring that currents in the reconnection
plane vanish, and hence, by Amp\`ere's Law, that the guide field (the
magnetic component parallel to the X-line) in the two plasmas be
equal.  However this choice has the peculiar consequence that there
are some magnetic field configurations for which reconnection cannot
occur because the reconnecting components of the field have the same
sign.  A further concern arises from the observation that when a
thermal pressure gradient exists at an X-line the guide field must
have spatial variations if the system is to be in total pressure
balance.  Since there is no a priori reason for assuming thermal
pressure gradients vanish at X-lines this calls the primary motivation
for Sonnerup's choice into question.

We propose a different criterion: reconnection occurs in the plane in
which the outflow speed from the X-line (given by an
appropriately-defined Alfv\'en speed) is maximized.  With this choice
reconnection can occur between any plasmas in which the magnetic
fields are not exactly parallel.  Reassuringly, the orientation of the
X-line also reduces to the expected result in symmetric cases.

\section{Definition of Coordinates}

It is particularly important for this problem to define the
coordinates carefully.  Consider two regions of plasma each with
number density $n_j$, temperature $T_j$, and magnetic field
$\mathbf{B}_j$, where $j =1,2$.  Assume that the two regions are
separated by a planar discontinuity through which no magnetic field
passes and define a coordinate system in which the $x$ and $z$ axes
lie in the discontinuity plane and the $y$ axis is perpendicular to
it.

Without any further constraint the X-line could, in principle, point
in any direction in the $x-z$ plane.  Each different X-line
orientation implies a different reconnection plane with different
components of the field reconnecting and a different reconnection
rate.  We want to find the X-line orientation for which reconnection
is fastest.  To do so it is most convenient not to consider a fixed
coordinate system in space but rather to define our coordinates with
respect to the direction of the reconnection X-line (the $z$
axis) and the plane of reconnection (the $x-y$ plane) and to rotate
the fields about the $y$ axis.  This rotation intermixes the guide,
$z$, and reconnecting, $x$, components of the fields and changes the
reconnection rate.  The GSM equivalents of our coordinates at the
magnetopause are $(x,y,z) \rightarrow (z,x,y)_{\text{GSM}}$.

Without loss of generality we specify the orientations of the fields
by defining $\theta$ to be the angle between the fields on either
side of the discontinuity (also called the shear angle) and $\alpha$
as the angle $\mathbf{B}_1$ makes with the $z$ axis (see Figure 1).
To make the problem well-defined we limit the ranges of the angles:
$0\leq \theta \leq \pi$ and $0 \leq
\alpha \leq \theta$.

The unknown parameter is $\alpha$ and varying $\alpha$ at fixed
$\theta$ changes the relative orientations of the fields with
respect to the X-line.  According to Sonnerup's argument $\alpha$ is
the solution of the equation $B_1 \cos\alpha =
B_2\cos(\theta-\alpha)$.  We claim that the proper choice is instead
the $\alpha$ that maximizes the outflow speed and the rate of
reconnection.

As an example, consider a system with $\theta = \pi$, $\mathbf{B}_1 =
-\mathbf{B}_2$, $n_1= n_2$, and $T_1 = T_2$.  These parameters
describe anti-parallel reconnection and symmetry suggests that $\alpha
= \pi/2 = \theta/2$.  Adding a constant guide field will change
$\theta$ but should keep $\alpha = \theta/2$.  Sonnerup's criterion
gives the expected results in these cases and ours, as will be seen,
does as well.  For other parameters, however, the two differ.

\begin{figure}
\noindent\includegraphics[width=3.0in]{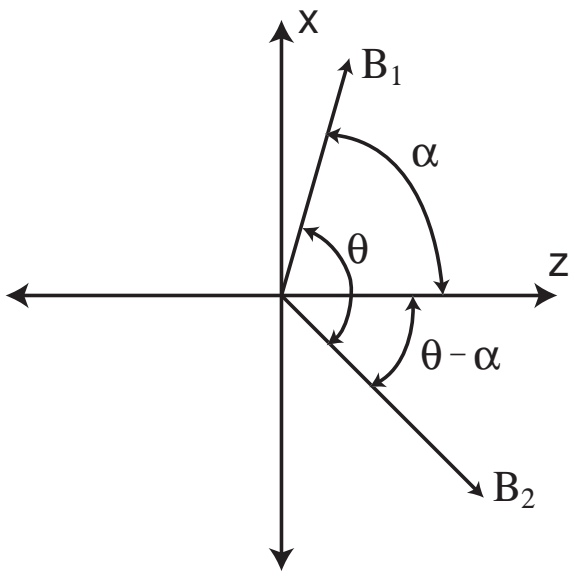}
\caption{\label{angles} Definition of the coordinate system.  The
plasmas meet in the plane shown, with one extending out of the page
and the other into the page.  The shear angle $\theta$ and the
directions of the $x$ and $z$ axes are fixed; $\alpha$ is unknown.}
\end{figure}

\section{Determining $\alpha$}

The rate at which magnetic field lines reconnect directly varies with
the speed at which they flow toward the X-line.  Continuity suggests
that the speed of this inflow is proportional to the speed of the
field lines' outflow, with a constant of proportionality that depends
on the detailed physics of the reconnection (e.g., the aspect ratio of
the diffusion region). For our purposes the details of the dependence
do not matter; the crucial point is that as the outflow speed
increases the reconnection rate does as well.

Since the outflow is driven by the motion of magnetic field lines it
must be related to some Alfv\'en speed; for symmetric anti-parallel
reconnection it is the speed calculated from the asymptotic field and
density.  Defining the appropriate outflow speed in the general case
is more complicated.  We find that it depends on the fields and
densities in both plasmas as well as the angles $\theta$ and $\alpha$.
Hence, the inflow speed and reconnection rate depend on these
quantities as well.

\subsection{Constructing the outflow speed}

\begin{figure}
\noindent\includegraphics[width=3.0in]{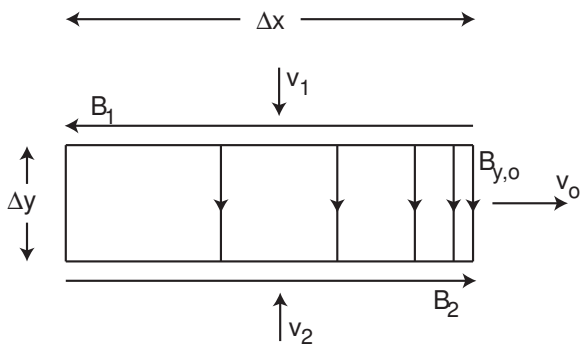}
\caption{\label{velocity} Cartoon of the current layer during
asymmetric reconnection.  The X-line is at the left of the box and the
asymptotic outflow is at the right.  All speeds and magnetic field
strengths are assumed to be positive.}
\end{figure}

Consider the situation shown in Figure \ref{velocity}.  The left side
of the box is the X-line where the plasma velocity and in-plane
magnetic field are assumed to vanish.  The plasma above the current
sheet has mass density $\rho_1 = m_1n_1$, where $m_1$ is the average
mass, and in-plane magnetic field $B_{x,1}$; below the current sheet
these values are $\rho_2$ and $B_{x,2}$. The plasmas flow into the
current sheet with speeds $v_1$ and $v_2$.  Within the current sheet
they mix in some proportion, resulting in a plasma of density
$\rho_0$, and accelerate downstream, dragged by the tension of the
reconnected magnetic field.  At the right-hand edge of the sheet the
plasma reaches its outflow speed $v_0$ and the in-plane field has a
magnitude $B_{y,0}$.  We assume the system is in a steady-state and
proceed to calculate $v_0$.

Applying conservation of mass to the box gives
\begin{equation}\label{mass}
\rho_0 v_o \Delta y = \rho_1 v_1\Delta x + \rho_2 v_2 \Delta x \, .
\end{equation}
The out-of-plane electric field $E_z$ is, according to Faraday's Law,
spatially constant in a 2-D steady-state system and, asymptotically,
must be given by the MHD result $\mathbf{E} =
-\mathbf{v}\boldsymbol{\times}\mathbf{B}/c$.  Equating the values at
the inflow and outflow edges of the current layer gives
\begin{equation}\label{ez}
v_1 B_{x,1} = v_2 B_{x,2} = v_0 B_{y,0} \, ;
\end{equation}
combining equations (\ref{mass}) and (\ref{ez}) yields an
expression for $\rho_0$
\begin{equation}\label{rho0}
\rho_0 = B_{y,0}\frac{\Delta x}{\Delta y}\left(\frac{\rho_1}{B_{x,1}} +
\frac{\rho_2}{B_{x,2}}\right)\, .
\end{equation}
Within the current layer the dominant terms in the $x$ component
of the momentum equation are advection and magnetic tension:
\begin{equation}
\rho_0 v_x \frac{\partial v_x}{\partial x} =
\frac{1}{4\pi}B_y\frac{\partial B_x}{\partial y} \, .
\end{equation}
We assume $B_x$ varies piecewise-linearly across the current layer and
rewrite this equation as
\begin{equation}\label{mom}
\frac{\partial}{\partial x}\, v_x^2 =
\frac{1}{4\pi}\frac{B_y}{\rho_0}\frac{B_{x,1}+B_{x,2}}{\Delta y} \, .
\end{equation}
After integrating with respect to $x$ along the current layer we have
\begin{equation}\label{notyet}
v_o^2 = \frac{1}{4\pi} \frac{B_{y,0}}{\rho_0} \frac{\Delta x}{\Delta y}
(B_{x,1} + B_{x,2}) \, ,
\end{equation}
where $\int B_y \,dx = B_{y,0} \Delta x$.  Combining equations
(\ref{rho0}) and (\ref{notyet}) gives the outflow speed
\begin{equation}\label{answer}
v_0^2 = \frac{B_{1x} +
B_{2x}}{4\pi}\left(\frac{\rho_1}{B_{1x}} +
\frac{\rho_2}{B_{2x}}\right)^{-1} .
\end{equation}
Equation (\ref{answer}) exhibits the necessary symmetry between the
two sides, reduces to the usual result, $v_0^2 = B_x^2/4\pi\rho$, when
$\rho_1 = \rho_2$ and $B_{1x} = B_{2x}$, and goes to zero, as
expected, when either density is large or either field vanishes.  This
result was independently derived in a slightly different context by
\cite{cassak07a}.

In terms of the angles defined in Figure \ref{angles} the outflow
speed is
\begin{multline}\label{parteq}
v_0^2 = \frac{B_1\sin\alpha +
B_2\sin(\theta-\alpha)}{4\pi} \\ \times
\left(\frac{\rho_1}{B_1\sin\alpha} +
\frac{\rho_2}{B_2\sin(\theta-\alpha)}\right)^{-1} \,.
\end{multline}
According to our previous argument the condition $\partial
v_0^2/\partial \alpha = 0$ defines the orientation of the X-line.

\subsection{Maximal Value}

Although the operations required to find an expression for $\alpha$
are straightforward, the actual calculations are a bit tedious.
Before presenting the result, we make some observations
\begin{enumerate}
\item $v_0^2(\alpha = 0) = v_0^2(\alpha = \theta) = 0$.  Since $v_0^2
\geq 0$ the implication is that $v_0^2$ has at least one maximum in the
range $0\leq \alpha \leq \theta$.  We strongly suspect, but have not
been able to prove, that there is only one maximum.
\item $B_1$, $B_2$, $\rho_1$, and $\rho_2$ are independent
variables but will only enter the result through the two dimensionless
ratios $b = B_2/B_1$ and $r = \rho_2/\rho_1$.  Hence $\alpha$ is a
function of only three parameters: $\theta$, $b$, and $r$.
\end{enumerate}

The solution for $\alpha$ is the root of the equation
\begin{multline}\label{maxcond}
0 = r\sin^2\alpha[\sin(\theta-2\alpha) - b\sin(2\theta-2\alpha)] \\ +
b\sin^2(\theta-\alpha)[\sin 2\alpha + b\sin(\theta-2\alpha)]
\end{multline}
subject to the constraint $0\leq \alpha \leq \theta$.  By defining
$\phi = \theta/2 - \alpha$, $\phi_+ = \theta/2 + \phi$, and $\phi_- =
\theta/2 -\phi$ equation (\ref{maxcond}) can be written in the
symmetric form
\begin{multline}
0 = r\sin^2\phi_-[\sin 2\phi - b\sin 2\phi_+] \\ +
b\sin^2\phi_+[b\sin 2\phi + \sin 2\phi_-]\, .
\end{multline}

Although equation (\ref{maxcond}) must, in general, be numerically
solved for $\alpha$, exact solutions are possible in some special cases
\begin{enumerate}
\item $\theta = \pi$ (anti-parallel reconnection).  In this case
$\alpha = \theta/2 = \pi/2$, independent of the values of $b$ and $r$.
\item $r = 1$ ($\rho_1 = \rho_2$).  Regardless of $b$ the maximal value
occurs for $\alpha = \theta/2$.  
\item $b \gg |1 - 1/r|$ or $b \ll |1/(1-r)|$.  Again the result is
$\alpha = \theta/2$.  The two limits are complementary in the sense that
the system is symmetric under the substitutions $b \rightarrow 1/b_*$,
$r \rightarrow 1/r_*$, $\alpha
\rightarrow \theta-\alpha_*$.
\end{enumerate}
The last example suggests that $\alpha = \theta/2$ is a good
approximation to the exact solution of equation (\ref{maxcond})
whenever the density ratio is not too much different from $1$.
Numerical trials bear this out, as can be seen in Figure 3 which shows
results for $b=2, r=0.5$.

\begin{figure}
\noindent\includegraphics[width=3.0in]{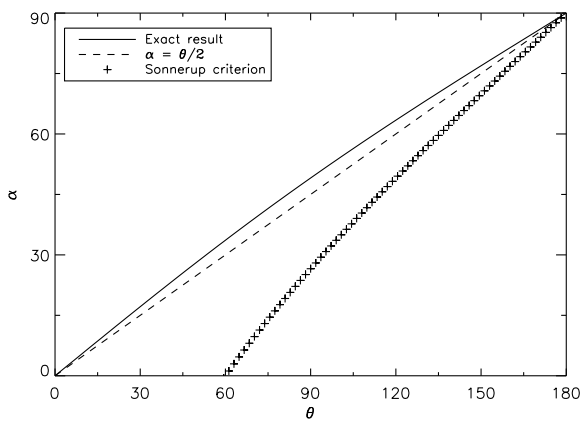}
\caption{\label{results} Plot of $\alpha$ versus the shear angle
$\theta$ for the case $b=2$, $r=0.5$.  Note that for Sonnerup's
solution there is no reconnection when $\theta \leq 60^{\circ}$.}
\end{figure}

Interestingly, since the outflow speed does not directly depend on
the temperatures or average masses of the plasmas, neither does
$\alpha$ (or, equivalently, the X-line orientation).  There is an
indirect constraint, however, because the system must also be in total
pressure balance,
\begin{equation}\label{pressbalance}
n_1T_1 + \frac{B_1^2}{8\pi} = n_2T_2 + \frac{B_2^2}{8\pi} \, ,
\end{equation}
if our assumption of steady-state reconnection is to
be valid.
If the temperature and the average mass are equal in the reconnecting
plasmas then equation (\ref{pressbalance}) relates $b$ and $r$ to the
plasma $\beta$
\begin{alignat}{2}
\beta_1 &= \frac{b^2-1}{1-r} & \qquad \beta_2 &=
\frac{r}{b^2}\frac{b^2-1}{1-r} \\
b^2 &= \frac{1+\beta_1}{1+\beta_2} & \qquad r &=
\frac{1+\beta_1^{-1}}{1+\beta_2^{-1}} \,.
\end{alignat}
If desired the condition of equation (\ref{maxcond}) can be re-written
in terms of $\beta_1$ and $\beta_2$.

\section{Discussion}

Establishing the system's orientation is an important part of the
interpretation of spacecraft observations.  Beginning with the basic
magnetic field data the well-known technique of minimum variance
analysis determines the direction normal to the current sheet (the $y$
axis in our coordinates).  Determining the direction of the X-line,
either through Sonnerup's criterion (see, for example,
\cite{phan06a}) or through equation (\ref{maxcond}), fixes the
geometry of the reconnection, provided only that the system has weak
variations along the direction of the X-line.  This information is
particularly important for those measurements that are to be compared
to theoretical models and simulations of reconnection, as will be the
case for the upcoming Magnetospheric Multiscale Mission.

Our proposed criterion can be checked with numerical simulations.
Since our argument does not depend on the detailed physics of the
reconnection, only that the reconnection rate varies with the Alfv\'en
speed, even MHD codes that do not correctly describe fast reconnection
should suffice.  However such simulations must take care not to impose
a reconnection plane a priori by, for example, not being fully
three-dimensional.

We emphasize that although we have attempted to calculate the
direction of the dominant reconnection X-line in a general current
layer in this paper, there are several possibly important effects that
have been neglected.  First, we cannot exclude the possibility that
reconnection may proceed simultaneously at different surfaces and
that, as a consequence, the current layer might become fully turbulent
\citep{galeev86a}.  Second, effects that preferentially suppress
reconnection for some X-line orientations are a possible complication
that we have ignored. \cite{swisdak03a} showed that a thermal pressure
gradient across the current layer drives diamagnetic drifts that
convect the X-line. As the drift speed approaches the Alfv\'en speed
the reconnection can be completely suppressed.  Since the magnitude of
the drift varies with the angle $\alpha$, the X-line orientation in
such systems may be determined by a trade-off between maximizing the
outflow Alfv\'en speed and minimizing the diamagnetic drift.  Other
effects, e.g., shear flows in the reconnecting plasmas, could have
similar consequences.

Finally, equation (\ref{maxcond}) determines the local
orientation of the X-line based on the parameters of the reconnecting
plasmas.  But what happens at, for instance, the magnetopause
where the shear angle can vary with location due to the combined
effects of the dipole tilt of the terrestrial field, the direction of
the interplanetary magnetic field, and the curvature of the interface?
Both the orientation of the X-line and the reconnection rate will then
vary with location with unknown effects on the global configuration of
the reconnection.  One possibility is that local maxima in the
reconnection rate will seed vigorously growing X-lines that propagate
outwards \citep{huba02b}, perhaps occasionally shifting directions to
merge with other reconnecting regions.  Depending on the external
conditions and length of time the system remains in a steady-state it
may have either a few or many simultaneously reconnecting X-lines.

\bibliographystyle{agu04}
\bibliography{paper}

\end{article}

\end{document}